\documentstyle[twoside, amssymb]{article}
\input amssym.def
\def\be{\begin{equation}}
\def\ee{\end{equation}}
\def\bea{\begin{eqnarray}}
\def\eea{\end{eqnarray}}
\def\bean{\begin{eqnarray*}}
\def\eean{\end{eqnarray*}}
\def\ba{\begin{abstract}}
\def\ea{\end{abstract}}
\def\bt{\begin{theorem}}
\def\et{\end{theorem}}
\def\bl{\begin{lemma}}
\def\el{\end{lemma}}
\def\br{\begin{remark}}
\def\er{\end{remark}}
\def\bc{\begin{corollary}}
\def\ec{\end{corollary}}
\def\bd{\begin{definition}}
\def\ed{\end{definition}}

\begin{document}
\baselineskip=20pt
\title{Unified Expressions Of All Integral Variational Principles}
\author{Yong-Chang Huang$^{1,4}$ Xi-Guo Lee$^2$ M. X. Shao$^3$ \\
\\
$^1$Institute of Theoretical Physics, College of Applied Sciences, \\
Beijing University of Technology, Beijing 100022, P. R. China\\
$^2$Institute of Modern Physics, Chinese Academy of Sciences, \\
P.O.Box 31, Lanzhou, 730000, China\\
$^3$Department of Physics, Beijing Normal University, \\
Beijing 100875, P. R. China\\
$^4$CCAST (World Lab.), P. O. Box 8730, Beijing 100080, P. R. China
}
\maketitle

\begin{abstract}
In terms of the quantitative causal principle, this paper obtains a general
variational principle, gives unified expressions of the general, Hamilton,
Voss, H\"{o}lder, Maupertuis-Lagrange variational principles of integral
style, the invariant quantities of the general, Voss, H\"{o}lder,
Maupertuis-Lagrange variational principles are given, finally the Noether
conservation charges of the general, Voss, H\"{o}lder, Maupertuis-Lagrange
variational principles are deduced, and the intrinsic relations among the
invariant quantities and the Noether conservation charges of the all
integral variational principles are achieved.
\end{abstract}

\section{Introduction}
Physical laws may be expressed by variational principle not only by
differential formulas, in fact, differential formulas can be derived from
variational principle\cite{des}, and there are the variational principles of
differential and integral styles\cite{cb,par}. Hilbert outlined 23 major
mathematical problems to be studied in the coming century, variational
problem is one of the problems. And Hilbert's address is still important and
should be researched by anyone interested in pursuing research in
mathematics and physics\cite{hilbe,hilb}.

The unified expressions of the all differential variational principles is
presented by the quantitative causal principle (QCP) derived from the
no-loss-no-gain principle in Ref.\cite{hdf}. Using the no-loss-no-gain
homeomorphic map transformation satisfying QCP, Ref.\cite{hcon} solves the
problem of the non-perfect properties of the Volterra process, gains the
exact strain tensor formulas in condensed matter theory. QCP reflects that
everything in the universe can not have cause without result or have result
without cause, and any serious science is always quantitatively expressed by
some equations, changes of some quantities in the equations must result in
the changes of the other quantities in the equations so that the right hand
sides of the equations keep no-loss-no-gain, namely, keep a kind of
invariant symmetry (i.e., maintain zero invariant symmetry ), therefore QCP
is general principle in physics, and QCP has the widest applications in both
physics and mathematics. It can be seen what this paper is a development of
Refs.\cite{hilbe,hilb,hdf,hsqu,cur}'s researches about the relationships of
causal principle, symmetry and variational principle. In statistical theory,
statistical quantitative causal relations in expressions of equations are
also satisfied, thus, definite results may be deduced\cite{hstat}.

In this paper, Sect. II gives the unified expressions of all integral
variational principles, Sect. III shows up conservation quantities of all
integral variational principles, Sect. IV represents Noether conservation
charges of all integral variational principles, the last Sect. is summary
and conclusions.

\section{Unified Expressions Of All Integral Variational
Principles}

Ref.\cite{hdf} rigorously gives the expression of QCP derived from the
no-loss-no-gain principle. In fact, in physics, the quantitative actions
(cause) of some quantities must lead to the corresponding equal effects
(result), which is just a expression of the no-loss-no-gain principle\cite
{hdf,hcon,hsqu}, therefore, we obtain\cite{hdf}

\emph{Quantitative Causal Principle: In a general physical system, how much
loses (cause), there must be, how much gains (result), or, quantitative
actions (cause) of some quantities must result in the corresponding equal
effects (result).}

The principle, then, may be concretely expressed as
\begin{equation}
DA-CA=0
\end{equation}
Eq.(2.1)'s physical meaning is that real physical result coming from any
operator's set D acting on A must result in appearance of set C acting on A
so that Eq.(2.1)'s right hand side seriously keeps no-loss-no-gain, namely,
making DA equate to CA, the whole process satisfies QCP with no-loss-no-gain
characteristic property. When A is a action, C is identity, and D is group
operator of infinitesimal continuous transformation, the physical process is
that the whole system with group D symmetry satisfies QCP, that is, the
action A takes limit value under the transformation of group D
\begin{equation}
\triangle A=A^{\prime }-A=\int_{t_1^{\prime }}^{t_2^{\prime
}}L^{\prime }(q^{\prime },\stackrel{.}{q^{\prime }},t^{\prime
})dt^{\prime }-\int_{t_1}^{t_2}L(q,\stackrel{.}{q},t)dt=0
\end{equation}
where $DA=A^{\prime }$, and the general infinitesimal transformations are%
\cite{dju}

\begin{equation}
t'=t+\triangle t,\mbox{ }q_i^{\ '(r)}=q_i^{\ ' (r)}+\triangle
q_i^{(r)} , (r=0,1)
\end{equation}

Not losing generality, let us define
\begin{equation}
L{' }(q',\stackrel{.}{q}',t')=L(q',%
\stackrel{.}{q}',t')+f(q',\stackrel{.}{q}',t')
\end{equation}
where f is a smooth function. It, thus, follows that
\begin{equation}
\triangle A=\int_{t_1}^{t_2}[f(q^{\prime },\stackrel{.}{q}^{\prime
},t^{\prime })+\triangle (f+L)+(f+L)\frac{d\Delta t}{dt}]dt
\end{equation}

Making Eq.(2.5) in order, we have

\begin{equation}
\triangle A=\int_{t_1}^{t_2}\{f(q,\dot{.}{q},t)+\sum{i}{\sum }[%
\frac{\partial (L+f)}{\partial q_i}-\frac d{dt}\frac{\partial (L+f)}{%
\partial \dot{.}{q}_i}]\delta q_i+\frac d{dt}[\sum{i}{\sum }%
\frac{\partial (L+f)}{\partial \dot{.}{q}_i}\delta q+(L+f)\Delta
t]\}dt
\end{equation}

Since Eq.(2.4) is a limit process, not losing generality, neglecting two-order
infinitesimal quantity, it follows that
\begin{equation}
f=f_0+\varepsilon _\sigma \frac{dg^\sigma }{dt},\mbox{ }\sigma
=1,2,\cdots ,m
\end{equation}
where $\varepsilon _\sigma $ ( $\sigma =1,2,\cdots ,m$) are
linearly independent infinitesimal parameters of Lie group D.

Substituting Eq.(2.7) into Eq.(2.6), neglecting two order infinitesimal
quantity, then Eq.(2.6) is simplified as

\begin{equation}
\triangle A=\int_{t_1}^{t_2}\{f_0(q,\dot{.}{q},t)+\sum{i}{\sum }[%
\frac{\partial (L+f_0)}{\partial q_i}-\frac d{dt}\frac{\partial (L+f_0)}{%
\partial \dot{.}{q}_i}]\delta q_i+\frac d{dt}[\sum{i}{\sum }%
\frac{\partial (L+f_0)}{\partial \dot{.}{q}_i}\delta
q_i+(L+f_0)\Delta t+g]\}dt
\end{equation}
in which g = $\varepsilon _\sigma g^\sigma ,\sigma =1,2,\cdots ,m.$ For the
third term in Eq.(2.8) taking the endpoint condition as zero, because $f_0(q,%
\dot{.}{q},t)$ does not contain one order infinitesimal quantity,
and using $\triangle A=0$ and the arbitration of one order
infinitesimal quantity $\delta q_i,$ we obtain f$_0=$ 0, Eq.(2.8) is
thus simplified as

\begin{equation}
\triangle A=\int_{t_1}^{t_2}\sum{i}{\{\sum }[\frac{\partial L}{%
\partial q_i}-\frac d{dt}\frac{\partial L}{\partial \dot{.}{q}_i}%
]\delta q_i+\frac d{dt}\sum{i}{(\sum }\frac{\partial L}{\partial
\dot{.}{q}_i}\delta q_i+L\Delta t+g)\}dt
\end{equation}

In the past, f$_0=$ 0 is well known widely to be the result keeping
Euler-Lagrange Equation invariant, now it is seen that under the condition
of Eq.(2.7), f$_0=0$ is not the result that maintains Euler-Lagrange equation
invariant, which is just the result satisfying QCP. This is the new result
that has not been obtained in the past.

Using Eq.(2.9) may give the unified expressions and the intrinsic relations of
all variational principles of integral style. In terms of the requirement
that integrating the second term in Eq.(2.9) vanishes, it follows that the
conditions of the new general variational principle are: for the fixed t$_1$
and t$_2$, $\delta q_i|_{t=t_1}=\delta q_i|_{t=t_2}=0$ , $\triangle
t|_{t=t_1}=\triangle t|_{t=t_2}=0$ and g(t$_1$) = g(t$_2$), and using
arbitration of $\delta q_i,$ we obtain Euler-Lagrange Equation

\begin{equation}
\frac{\partial L}{\partial q_i}-\frac d{dt}\frac{\partial
L}{\partial \dot{.}{q_i}}=0
\end{equation}

When taking equal time $\triangle $t = 0 in Eq.(2.9), we obtain Hamilton
variational principle

\begin{equation}
\delta A=\int_{t_1}^{t_2}\{\frac d{dt}[\sum{i}{\sum }\frac{\partial L%
}{\partial \dot{.}{q}_i}\delta q_i+g]+\sum{i}{\sum }[\frac{%
\partial L}{\partial q_i}-\frac d{dt}\frac{\partial L}{\partial \dot{.}{%
q_i}}]\delta q_i\}dt
\end{equation}

Using the requirement that integrating the first term in Eq.(2.11) is zero, it
follows that the conditions of Hamilton variational principle are: for the
fixed t$_1$ and t$_2$, $\delta q_i|_{t=t_1}=\delta q_i|_{t=t_2}=0$ and g(t$%
_1 $) = g(t$_2$), and looking from E$^q$ space, which is fixed endpoint
variational. Using arbitration of $\delta q_i,$we get Euler-Lagrange
Equation (2.10). Due to

\begin{equation}
\triangle q_i^{(r)}=\delta q_i^{(r)}(t^{\prime })+q_i^{(r+1)}\triangle t,%
\mbox{ }r=0,1
\end{equation}

Eq.(2.9) can be rewritten as

\begin{equation}
\triangle A=\int_{t_1}^{t_2}\{\sum{i}{\sum }(\frac{\partial L}{%
\partial q_i}-\frac d{dt}\frac{\partial L}{\partial \dot{.}{q}_i}%
)(\triangle _pq_i-\dot{.}{q_i}\triangle t)+\frac d{dt}[\sum{i}{%
\sum }(\frac{\partial L}{\partial \dot{.}{q}_i})(\triangle _pq_i-%
\dot{.}{q_i}\triangle t)+L\triangle t+g]\}dt
\end{equation}

Because the requirement that integrating the second term in Eq.(2.13)
vanishes, we obtain that the conditions of Voss variational principle are:
unequal time variational, $\triangle t|_{t=t_1}=\triangle t|_{t=t_2}=0,$ $%
\triangle _pq_i|_{t=t_1}=\triangle _pq_i|_{t=t_2}=0$ and g(t$_1$) = g(t$_2$%
). Because $\triangle _pq_i-\dot{.}{q_i}\triangle t$
$(i=1,2,\cdots ,n.) $ construct virtual variational of the system,
we similarly get the Euler-Lagrange Equation (10) from Eq.(2.13).
When inserting Eq.(2.12) into Eq.(2.9), and using

\begin{equation}
\frac d{dt}(\sum{i}{\sum }\frac{\partial L}{\partial \dot{.}{q}_i%
}\dot{.}{q_i}-L)=-\sum{i}{\sum }(\frac{\partial L}{\partial q_i}%
-\frac d{dt}\frac{\partial L}{\partial \dot{.}{q}_i})\dot{.}{q_i}-%
\frac{\partial L}{\partial t}
\end{equation}

\begin{equation}
\triangle A=\int_{t_1}^{t_2}(\triangle L+L\frac{d\triangle t}{dt}+\frac{dg}{%
dt})dt
\end{equation}
we further make Eq.(2.9) in order, it follows that

\begin{equation}
\int_{t_1}^{t_2}\{\triangle L-\sum{i}{\sum }(\frac{\partial L}{%
\partial q_i}-\frac d{dt}\frac{\partial L}{\partial \dot{.}{q}_i}%
)\triangle q_i+\sum{i}{\sum }\frac{\partial L}{\partial \dot{.}{q%
}_i}\dot{.}{q}_i\frac{d\triangle t}{dt}-\frac{\partial L}{\partial t}%
\triangle t\}dt=\int_{t_1}^{t_2}\sum{i}{\sum }\frac d{dt}[(\frac{%
\partial L}{\partial \dot{.}{q}_i})\triangle q_i]dt
\end{equation}

Using the demand that keeps Eq.(2.16)'s right hand side no-loss-no-gain to
equate zero, we obtain that the conditions of H\"{o}lder variational
principle are: unequal time variational, $\triangle $t may be not to equate
to zero and $\triangle q_i|_{t=t_1}=\triangle q_i|_{t=t_2}=0.$

When the system satisfies the Euler-Lagrange Eq.(2.10), we get the simplified
expression of Eq.(2.16) of H\"{o}lder variational principle as follows

\begin{equation}
\int_{t_1}^{t_2}\{\triangle L+\sum{i}{\sum }[\frac{\partial L}{%
\partial \dot{.}{q}_i}\dot{.}{q}_i]\frac{d\triangle t}{dt}-\frac{%
\partial L}{\partial t}\triangle t\}dt=0
\end{equation}

It is seen that H\"{o}lder variational principle is originated from Eq.(2.9)
satisfying QCP yet, looking in E$^q$ space, the variational is the
variational of variant endpoint along time t axis.

When taking equal time variational in Eq.(2.17), it is simplified as Hamilton
variational principle\cite{ros}

\begin{equation}
\delta A=\int_{t_1}^{t_2}\delta Ldt=0
\end{equation}

It is looked that Hamilton variational principle not containing g
is yet a simplified expression of H\"{o}lder variational principle
under the condition of equal time variational. In expression (2.17)
of H\"{o}lder variational principle, as L = L($q,\dot{.}{q}$), one
has $\partial L/\partial t=0$. When there exists motion trajectory
integration\cite{ros}

\begin{equation}
H=T+V=\sum{i}{\sum }\frac{\partial L}{\partial \dot{.}{q}_i}%
\dot{.}{q_i}-L=const.
\end{equation}
it follows from Eq.(2.19) that

\begin{equation}
\triangle H=\triangle \sum{i}{\sum }(\frac{\partial L}{\partial
\dot{.}{q}_i}\dot{.}{q_i})-\triangle L=0
\end{equation}

Substituting $\partial L/\partial t=0$ and Eq.(20) into Eq.(2.17), we have

\begin{equation}
\int_{t_1}^{t_2}(\triangle (2T)+2T\frac{d\triangle
t}{dt})dt=\triangle \int_{t_1}^{t_2}(2T)dt=0
\end{equation}
where $\sum{i}{\sum }(\partial L/\partial \dot{.}{q}_i)\dot{%
}{q_i}=2T$ has been used.

Eq.(21) is just the expression of Maupertuis-Lagrange variational principle.
It is watched that the Maupertuis-Lagrange principle's conditions is
H\"{o}lder's, i.e, the endpoints are motional\cite{par}. Therefore, the
unified expressions of the all variational principles of integral style is
shown by means of QCP, and the other high order principles etc can be
analogously discussed.

\section{Conservation Quantities of All Integral Variational Principles}

Now we discuss their conservation quantities about the above all principles.

Taking endpoint condition $(\cdots )(t_1)=$ $(\cdots )(t_2)$ in Eq.(2.9), when
using arbitrations of t$_1,$ t$_2$ and t in given [t$_{01},$t$_{02}$] and $%
\delta q_i$, we obtain Euler-Lagrange Eq.(2.10) and the following general
conservation quantity of the general variational principle

\begin{equation}
\sum{i}{\sum }\frac{\partial L}{\partial \dot{.}{q}_i}\delta
q_i+L\triangle t+g=const
\end{equation}

Analogous to the discussions of Eq.(3.22), the conservation quantity of Voss
variational principle may be obtained. On the other hand, because Eq.(3.22) is
the general conservation quantity, the conservation quantity of Voss
variational principle can also be achieved by substituting $\delta q_i=$ $%
\triangle _pq_i-\dot{.}{q_i}\triangle t$ into the general
conservation expression (3.22) as follows

\begin{equation}
\sum{i}{\sum }\frac{\partial L}{\partial \dot{.}{q}_i}(\triangle
_pq_i-\dot{.}{q_i}\triangle t)+L\triangle t+g=const
\end{equation}

The general integral conservation quantity Eq.(3.22) is compared with the
differential conservation quantity\cite{hdf}

\begin{equation}
\sum{i}{\sum }\frac{\partial L}{\partial \dot{.}{q}_i}\delta
q_i+L\triangle t=const,\mbox{ }\sigma =1,2,\cdots ,m
\end{equation}
g is the more part than the differential style's. When taking equal time
variational in the general conservation expression (3.22), we obtain
conservation quantity of Hamilton variational principle

\begin{equation}
\sum{i}{\sum }\frac{\partial L}{\partial \dot{.}{q}_i}\delta
q_i+g=const.\mbox{ }
\end{equation}

In the similar reason, it follows from Eq.(2.16) that the conservation
quantity of H\"{o}lder variational principle
\begin{equation}
\sum{i}{\sum }\frac{\partial L}{\partial \dot{.}{q}_i}\triangle
q_i=const.\mbox{ }
\end{equation}

Since the expression (2.21) of Maupertuis-Lagrange variational principle is
just the result of H\"{o}lder variational principle under the constraint
conditions (2.19) and $\partial L/\partial t=0,$ thus the corresponding
conservation quantity is still Eq.(3.26). Substituting $L=2T-const.$ into
Eq.(3.26) we further deduces the conservation quantity of Maupertuis-Lagrange
variational principle as follows

\begin{equation}
\sum{i}{\sum }\frac{\partial 2T}{\partial \dot{.}{q}_i}\triangle
q_i=const.
\end{equation}

It is well known that it is usually neglected that the invariant quantities
of Voss, H\"{o}lder, Maupertuis-Lagrange variational principles, thus, the
invariant quantities of the new general, Voss, H\"{o}lder,
Maupertuis-Lagrange variational principles are first given, then
conservation quantities of all integral variational principles are first
deduced, and the intrinsic relations among the invariant quantities of the
all integral variational principles are first found.

\section{Noether Conservation Charges of All Integral Variational Principles}

Now we use the conservation quantities above to find the corresponding
Noether conservation charges of the systems with invariant properties of Lie
group D$_m.$ Let's consider the infinitesimal transformations of the
spacetime coordinates by Lie group D$_m$ as follows\cite{dju,liz}

\begin{equation}
t^{\prime }=t^{\prime }(q(t),\dot{.}{q}(t),t,\alpha )\doteq
t+\Delta t=t+\varepsilon _\sigma \tau ^\sigma \mbox{ }(\sigma
=1,2,\cdots ,m)
\end{equation}
\begin{equation}
q_i^{\prime \mbox{ }(r)}=q_i^{\prime \mbox{ }(r)}(q(t),\dot{.}{q}%
(t),t,\alpha )\doteq q_i^{(r)}+\Delta
q_i^{(r)}=q_i^{(r)}+\varepsilon _\sigma (\xi _i^\sigma
)^{(r)}\mbox{ , }(\mbox{ }r=0,1)
\end{equation}
where $\varepsilon _\sigma $ $(\sigma =1,2,\cdots ,m)$ are infinitesimal
parameters corresponding to $\alpha _\sigma ,$ $\alpha _\sigma $ $(\sigma
=1,2,\cdots ,m)$ are m linearly independent infinitesimal continuous
transformation parameters of Lie group D$_m,$ $\tau ^\sigma $ and $(\xi
_i^\sigma )^{(r)}$ are the infinitesimal generated functions as follows

\begin{equation}
\tau ^\sigma =\frac{\partial t^{\prime }(q(t),\dot{.}{q}(t),t,\alpha )}{%
\partial \alpha _\sigma }\mid _{\alpha =0}(\sigma =1,2,\cdots ,m)
\end{equation}

\begin{equation}
(\xi _i^\sigma )^{(r)}=\frac{\partial q_i^{\prime \mbox{ }(r)}(q(t),%
\dot{.}{q}(t),t,\alpha )}{\partial \alpha _\sigma }\mid _{\alpha
=0}(\sigma =1,2,\cdots ,m;r=0,1)
\end{equation}

Using Eqs.(2.7), (2.12), (3.22), (4.28-31) and $\varepsilon _\sigma ^{}$'s
arbitration, we obtain the Noether conservation charges of the general
variational principles as follows
\begin{equation}
\sum{i}{\{\sum }\frac{\partial L}{\partial \dot{.}{q}_i}((\xi
_i^\sigma )-\dot{.}{q}_i\tau ^\sigma )+L\tau ^\sigma +g^\sigma
\}(t)=const.\mbox{ }(\sigma =1,2,\cdots ,m)
\end{equation}

Making use of Eqs.(2.7), (2.12), (3.23), (4.28-31) and $\varepsilon _\sigma ^{}$'s
arbitration, we get the Noether conservation charges of Voss variational
principles is still Eq.(3.23) but substituting $(\xi _i^\sigma )$ with $(\xi
_i^\sigma )_p$.

Using Eqs.(2.7), (2.12), (3.25), (4.28-31) and $\varepsilon _\sigma ^{}$'s
arbitration or in the general Eq.(32) let $\Delta t=\tau ^\sigma $ = 0, we
obtain Noether conservation charges of Hamilton variational principle

\begin{equation}
\sum{i}{\sum }\frac{\partial L}{\partial \dot{.}{q}_i}(\xi
_i^\sigma )(t)+g(t)^\sigma =const.\mbox{ }(\sigma =1,2,\cdots ,m)
\end{equation}

Utilizing $\varepsilon _\sigma ^{}$'s arbitration or in the general Eq.(4.32)
let g = 0 and $\Delta t=\tau ^\sigma $ = 0, we obtain Noether conservation
charges of H\"{o}lder variational principle as follows

\begin{equation}
\sum{i}{\sum }\frac{\partial L}{\partial \dot{.}{q}_i}(\xi
_i^\sigma )(t)=const.\mbox{ }(\sigma =1,2,\cdots ,m)
\end{equation}

Let $L=2T-const.$ , we can obtain Noether conservation charges of
Maupertuis-Lagrange variational principle
\begin{equation}
\sum{i}{\sum }\frac{\partial 2T}{\partial \dot{.}{q}_i}(\xi
_i^\sigma )(t)=const.\mbox{ }(\sigma =1,2,\cdots ,m)
\end{equation}

We, thus, get the conclusion that Noether conservation charge of the general
variational principle is general, and the Noether conservation charges of
the other variational principles are the special examples of the general
variational principle under different conditions. And it can be seen from
the above researches that all variational principles of integral style
deduce the same Euler-Lagrange Eq.(2.10), and there may be the different
conservation quantities of the different systems, The former characterizes
that physical laws don't depend on different conditions, the latter
characterizes that physical manifestations of systems may be many kinds of
variations, because the invariant quantities are relative to the physical
observable quantities. Since the quantitative causal principle is more
general, its applications to high order Lagranges etc will be written in the
other papers.

\section{Summary and Conclusion}

In terms of the mathematical expression of QCP, this paper gives a general
variational principle, shows unified expressions of the general, Hamilton,
Voss, H\"{o}lder, Maupertuis-Lagrange variational principles of integral
style, finds the intrinsic relations among the different integral
variational principles, it is shown that under the condition of Eq.(2.7), f$%
_0=0$ is not the result that maintains Euler-Lagrange equation invariant,
which is just the result satisfying QCP. The invariant quantities of the
general, Voss, H\"{o}lder, Maupertuis-Lagrange variational principles are
given, and the intrinsic relations among the invariant quantities and the
Noether conservation charges of the all integral variational principles are
found. In fact, the above discussions and Ref.\cite{hdf} make the
expressions of the past scrappy numerous variational principles be unified
into the relative consistent system of the all variational principles in
terms of QCP, which is essential for researching the intrinsic relations
among the past scrappy numerous variational principles and their Noether
theorems and for further making their logic simplification and clearness.

ACKNOWLEDGMENTS:

This work was supported by Chinese Academy of Sciences Knowledge Innovation
Project (KJCX2-SW-No2), National Natural Science Foundations of China
(10435080, 10575123).

The authors are grateful to Prof. Z. P. Li for useful
discussion.

\end{document}